\documentclass[sigconf,9pt,nonacm]{acmart}
\setcopyright{none}
\usepackage[T1]{fontenc}

\usepackage{xcolor, graphicx} %

\usepackage{listings}
\usepackage{amsmath}
\usepackage{amsfonts}
\usepackage{amsthm}
\usepackage{isomath}
\usepackage{color}
\usepackage{tikz}
\usepackage{svg}
\usepackage{balance}
\usepackage{enumitem}
\usepackage{hyperref}

\usepackage{pifont}
\usepackage{threeparttable}
\usepackage{subcaption}
\usepackage[most]{tcolorbox}
\usepackage{mdframed}

\allowdisplaybreaks[4]

\usetikzlibrary{positioning,automata}

\newtheorem*{definition*}{Definition}

\newtheorem*{lemma*}{Lemma}

\newcommand{\ket}[1]{\vert #1 \rangle}

\title{RISC-Q: A Generator for Real-Time Quantum Control System-on-Chips Compatible with RISC-V }
\author{Junyi Liu, Yi Lee, Haowei Deng, Connor Clayton, Gengzhi Yang, Xiaodi Wu}
\affiliation{
  \institution{Joint Center for Quantum Information and Computer Science, University of Maryland, College Park}
  \country{}
}
 \vspace{-0.25in}

\date{}

\begin{document}

\begin{abstract}
Quantum computing imposes stringent requirements for the precise control of large-scale qubit systems, including, for example, microsecond-latency feedback and nanosecond-precision timing of gigahertz signals—demands that far exceed the capabilities of conventional real-time systems. The rapidly evolving and highly diverse nature of quantum control necessitates the development of specialized hardware accelerators. While a few custom real-time systems have been developed to meet the tight timing constraints of specific quantum platforms, they face major challenges in scaling and adapting to increasingly complex control demands—largely due to fragmented toolchains and limited support for design automation. To address these limitations, we present RISC-Q---an open-source flexible generator for Quantum Control System-on-Chip (QCSoC) designs, featuring a programming interface compatible with the RISC-V ecosystem. Developed using SpinalHDL, RISC-Q enables efficient automation of highly parameterized and modular QCSoC architectures, supporting agile and iterative development to meet the evolving demands of quantum control. We demonstrate that RISC-Q can replicate the performance of existing QCSoCs with significantly reduced development effort, facilitating efficient exploration of the hardware–software co-design space for rapid prototyping and customization.
\end{abstract}

\maketitle

\section{Introduction}
Quantum computing’s potential will be unlocked only when systems can scale reliably and efficiently.
Although today’s quantum computers have not yet reached practical utility, their rapid scaling is already intensifying demands on the quantum control stack. 
Quantum control systems, once built for lab prototypes, must now meet industrial demands—delivering far greater precision, timing accuracy, and scalability. As a result, they are no longer peripheral hardware but the critical infrastructure enabling practical, high-performance quantum computing.

The fundamental control requirements for quantum systems already exceed the capabilities of conventional real-time systems, necessitating customized Quantum Control System-on-Chip (QCSoC) solutions. 
Superconducting qubits require gigahertz control, nanosecond timing, and microsecond feedback, while atomic platforms bring their own challenges—ion shuttling, optical-tweezer control, atom rearrangement, and high-speed imaging—all of which require dedicated real-time hardware.

As quantum hardware advances, new control challenges continue to emerge. Mid-circuit measurement demands the control capability for real-time feedback (e.g.,~\cite{Oliver-feedback-NC}) and dynamic circuit (e.g.,~\cite{CarreraVazquez2024,PRXQuantum.5.030339}). Quantum error correction~\cite{RevModPhys.87.307} demands an integrated solution of syndrome extraction, decoding, and correction within tight coherence windows.
Scalable quantum control also requires resource-efficient architectures, coordinated QCSoCs, and tight GPU integration to enable advanced quantum–classical processing~\cite{caldwell2025platform}.

Existing solutions—whether open-source (e.g., ~\cite{xu2023qubic,stefanazzi2022qick,kasprowicz2020artiq,kulik2022latest}), proprietary (e.g., ~\cite{fu2019eqasm,gebauer2023qicells,guo2023hisep}), or commercial (e.g., Zurich Instruments, Quantum Machines, Qblox)—can handle basic control for small quantum systems but struggle to keep pace with rapidly scaling requirements.
Two systemic issues limit their evolution: (1) \textbf{fragmented toolchains}, where hardware-specific architectures and APIs create vendor lock-in and hinder cross-platform development; and (2) \textbf{limited design automation}, which forces researchers to hand-craft low-level control architectures instead of relying on high-level automated tools.

Inspired by agile hardware development in machine-learning accelerators~\cite{genc2021gemmini,AHA-CGRA-TECS}, we introduce RISC-Q—the first open-source QCSoC generator with a RISC-V–compatible programming interface.
We expect extensive design exploration will be needed to identify optimal control architectures for diverse quantum platforms. RISC-Q enables this by automating highly parameterized, modular QCSoC designs, supporting rapid, iterative hardware–software co-design.
By providing a unified, extensible platform, RISC-Q also promotes interoperability and community-driven innovation in quantum control system design.

\begin{figure*}[t]
  \centering
  \begin{subfigure}[t]{0.25\linewidth}
    \includegraphics[width=\linewidth]{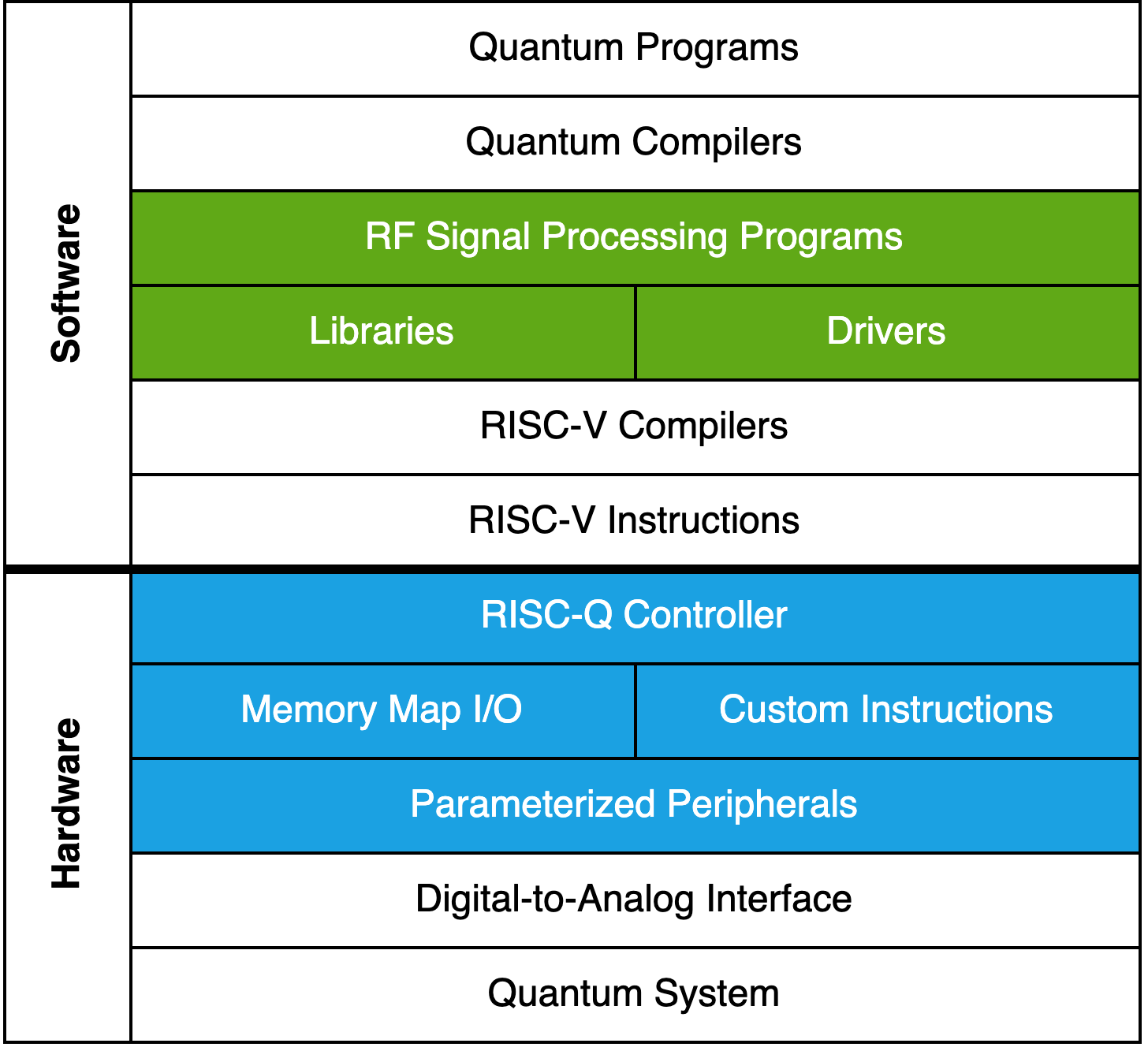}
    \caption{}
    \label{fig-intf}
  \end{subfigure}
  \hfill
  \begin{subfigure}[t]{0.7\linewidth}
    \includegraphics[width=\linewidth]{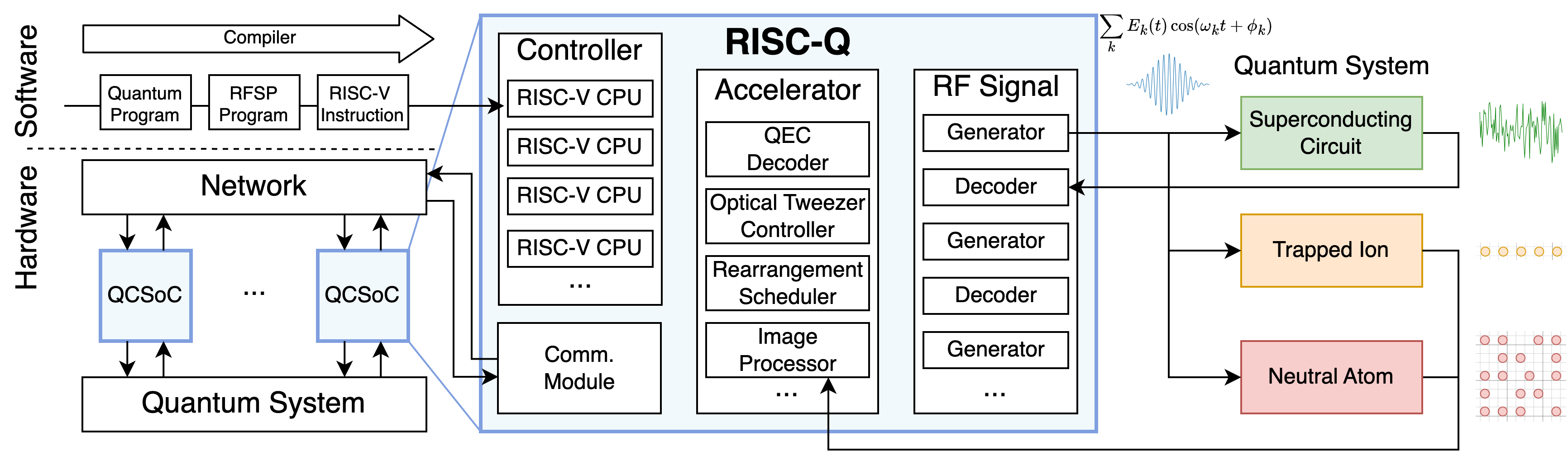}
    \caption{}
    \label{fig-arch}
  \end{subfigure}
  \caption{(a) Software and hardware stack of RISC-Q. Highlight boxes indicate components implemented within the RISC-Q framework. (b) A Schematic Overview of the RISC-Q hardware architecture for Quantum Control System-on-Chip (QCSoC). }
\end{figure*}

Specifically, RISC-Q is a modular and flexible QCSoC generator implemented in SpinalHDL~\cite{spinalhdl} with integrated RISC-V controllers, which allows RISC-V-compatible programming interface in Fig.~\ref{fig-intf} for enhanced scalability and interoperability. RISC-Q development adopts a general model of quantum control systems in Fig.~\ref{fig-arch}, representing diverse platforms using digital logic for control and analog RF components for interfacing with quantum hardware.

We present a fully functional QCSoC for superconducting quantum computers, built with RISC-Q and deployed on a Xilinx ZCU216 RFSoC board.
Our RISC-Q–designed QCSoC matches or exceeds state-of-the-art QICK~\cite{xu2023qubic} and QubiC~\cite{stefanazzi2022qick} systems across all key metrics (Table~\ref{table-performance}), including maximum frequency, qubit capacity and number of output lines/tones per board,  feedback and network latency, and system jitter, while significantly reducing both development effort and codebase size (by $\sim$75\%).

Furthermore, we demonstrate advanced quantum control capabilities enabled by the RISC-V programming interface and modular SoC architecture of RISC-Q. Specifically, we showcase co-designed conditional gates triggered by mid-circuit measurements, on-chip execution of calibration protocols to reduce communication overhead and accelerate feedback, as well as a seamless integration of quantum control and quantum error correction enabled by RISC-Q. 

In brief, this work makes the following contributions: 
\begin{enumerate}[leftmargin=12pt]
    \item We design RISC-Q, an open-source, modular QCSoC generator with a RISC-V–compatible interface, supporting flexible parameterization and peripheral integration. %
    \item Using RISC-Q, we implement a fully functional QCSoC deployable for superconducting qubit control, surpassing state-of-the-art performance while drastically reducing development effort. 
    \item We demonstrate that RISC-Q enables efficient co-design of conditional branching, supports direct on-chip execution of calibration protocols, and provides seamless integration between control and QEC systems.
\end{enumerate}

\vspace{-2mm}
\section{Modeling Quantum Control Systems} \label{sec:model}

\textbf{Physical Requirements:}
Several leading quantum computing platforms—such as superconducting circuits, neutral atoms, and trapped ions—rely on radio-frequency (RF) analog signals for control, despite differences in their underlying physics. These platforms impose far more stringent real-time requirements than classical systems. In superconducting qubits, for example, control signals must operate in the 3–6 GHz range, with precise phase coherence essential because the signal phase directly sets the gate rotation axis~\cite{krantz2019quantum}.
Prior research~\cite{xu2023qubic,stefanazzi2022qick,kasprowicz2020artiq} has demonstrated that high-performance digital control systems can generate analog signals that meet the demanding requirements of quantum systems.

Fig.~\ref{fig-arch} shows a representative quantum control system, with components grouped by function: \emph{RF signal processors, controllers, custom accelerators}, and \emph{communication modules}. The combination of architecture, implementation, and parameters creates a vast design space. Efficient exploration is crucial to meet stringent real-time constraints, develop new functionalities,  while staying within hardware limits, which motivates the development of RISC-Q. 

\vspace{-2mm}
\subsection{RF Signal Processor}

Quantum systems are driven by RF signals, which require dedicated modules for generation and decoding.

\vspace{-1mm}
\subsubsection{\textbf{RF Signal Generator}}
The RF signal generator produces the signals that drive quantum systems. In digital control systems, these signals are generated by outputting discrete samples to a DAC (Digital-to-Analog Converter), which transforms them into analog signals.
The RF signal generator is the most real-time critical component. The phase $\phi_k$ of a driving signal directly affects the resulting quantum gate. With carrier frequencies $\omega_k$ in the MHz-GHz range, even a single-cycle timing error can result in significant phase deviation. 
For superconducting qubits, the high carrier frequencies (several GHz) demand extremely precise timing. 
Neutral atom and trapped ion systems typically demand multi-tone signals at lower frequencies but with greater spectral complexity~\cite{young2020half}.

\vspace{-1mm}
\subsubsection{\textbf{RF Signal Decoder}}
In superconducting systems, qubit measurement involves sending a readout RF signal and extracting the phase from the feedback signal. Since outcomes are needed for real-time tasks such as quantum error correction and dynamic circuits, decoding latency is critical. Noise in the readout also affects accuracy, making precise decoding essential. The standard approach demodulates the I/Q components of the feedback signal and applies low-pass filtering to extract the phase~\cite{krantz2019quantum}.

\vspace{-2mm}
\subsection{Controller}
The controller provides a programmable interface to the quantum control system and must support the diverse signal sequences and feedback logic required by different experimental protocols. General-purpose CPUs are commonly used to enable flexible, software-defined control without hardware redesign, making them a natural choice. A stable, flexible programming interface built on a well-tested instruction set architecture (ISA) is crucial, as it ensures a mature software ecosystem and keeps control software reusable across hardware revisions and collaborative development cycles.

Beyond software flexibility, the controller must integrate heterogeneous components such as RF processors and custom accelerators. It must also satisfy strict real-time constraints: nanosecond-level timing for phase-sensitive operations, hundreds-of-megahertz clock rates for RF processing, and microsecond-scale feedback latency to support real-time protocols like quantum error correction.

\vspace{-2mm}
\subsection{Custom Accelerator}
Although general-purpose controllers provide programmability, they often struggle to meet real-time demands for computationally intensive quantum control tasks. To overcome this limitation, custom accelerators are introduced to handle the heavy computations required by advanced quantum control protocols.

\emph{QEC decoders} are a prime example of necessary hardware accelerators: they must process measurement outcomes and identify errors in real time. Practical decoders often rely on algorithms such as minimum-weight perfect matching, belief propagation, or neural-network inference—methods that are computationally intensive and frequently exceed the strict latency constraints of quantum systems. As a result, specialized hardware acceleration is essential to meet real-time demands and ensure timely error correction~\cite{google2024qec,caune2024demonstrating,wu2025micro}.

Neutral atom and trapped-ion platforms could rely on additional fast, hardware-accelerated control modules to meet stringent real-time requirements. Optical sensors generate raw images that must be rapidly processed to extract qubit positions and measurement outcomes, motivating dedicated \emph{image-processing} hardware to reduce feedback latency~\cite{wang2023accelerating}. Neutral atom systems also depend on \emph{optical tweezer controllers}, which convert desired atom positions into RF drive parameters for AOD-based beam steering~\cite{young2020half}. In addition, probabilistic loading necessitates on-the-fly rearrangement of atomic arrays; conventional PC-based scheduling introduces prohibitive round-trip delays, prompting recent proposals for hardware-accelerated, on-chip \emph{rearrangement schedulers} to eliminate these latency bottlenecks~\cite{wang2023accelerating,guo2024design}.

\vspace{-2mm}
\subsection{Communication Module}
Practical quantum computing involves controlling thousands to millions of qubits, which necessitates a distributed architecture \cite{fruitwala2024distributed}. Each node in the system requires a communication module to \emph{exchange} data and \emph{synchronize} operations.

Communication tasks include sharing measurement outcomes, delivering gate instructions, and synchronizing clocks across controllers. For instance, QEC protocols require real-time sharing of measurement results to compute error syndromes and apply corrections \cite{caune2024demonstrating}. Achieving high fidelity thus requires extremely low-latency communication.
Clock synchronization is also crucial: phase coherence between controllers depends on precise alignment in time. Therefore, the communication module must support high-precision synchronization protocols.

\vspace{-2mm}
\section{Design of RISC-Q}
\label{sec-arch}

RISC-Q is a modular generator for RISC-V–compatible quantum control SoC designs. Built with SpinalHDL~\cite{spinalhdl}, it provides parameterized hardware templates and high-level abstractions that decouple software and hardware development, enabling reusable components and facilitating collaborative development of QCSoCs.
\vspace{-2mm}

\subsection{Programming Interface}
RISC-Q adopts the RISC-V ISA as the programming interface for quantum control, leveraging open-source toolchains and established development workflows. This choice unifies the software interface across hardware platforms and avoids the fragmented ecosystems created by custom ISAs~\cite{kasprowicz2020artiq,stefanazzi2022qick,xu2023qubic}.

Quantum algorithms are programmed as RF Signal Processing (RFSP) process on quantum  hardware. Example RFSP programs can be obtained by replacing the gates and measurements in OpenQASM~\cite{cross2022openqasm} programs with functions that manipulate the RF signal processors to perform the corresponding operations. Fig. \ref{fig-qasm-c-rfsp} shows a simple example in which an \texttt{x} gate becomes an explicit sequence of RF-configuration operations in C.

\begin{figure}[h]
  \centering
    \begin{mdframed}[ innerleftmargin=2pt, innerrightmargin=2pt, innertopmargin=0pt, innerbottommargin=0pt ]
    \begin{lstlisting}[language={},basicstyle=\ttfamily\scriptsize]
// OpenQASM program for an x gate
qubit[8] q[0];
x q[0];
    \end{lstlisting}
    \end{mdframed}

    \begin{mdframed}[ innerleftmargin=2pt, innerrightmargin=2pt, innertopmargin=0pt, innerbottommargin=0pt ]
    \begin{lstlisting}[language=C,basicstyle=\ttfamily\scriptsize]
// RFSP program in C that drives the signal generator for x gate
x_gate(0, TIME_NS(100)); // apply x gate on qubit 0 at 100 ns

inline void x_gate(int id, int start_time) {
  *(START_TIME)           = start_time;
  *(PULSE_FREQ_ADDR(id))  = GATE_X_FREQ(id);
  *(PULSE_PHASE_ADDR(id)) = GATE_X_PHASE(id);
  *(PULSE_AMP_ADDR(id))   = GATE_X_AMP(id);
  *(PULSE_ENV_ADDR(id))   = GATE_X_ENV(id);
  *(PULSE_DUR_ADDR(id))   = GATE_X_DUR(id); }
    \end{lstlisting}
    \end{mdframed}
  \caption{Side-by-side comparison of an \texttt{x} gate written in OpenQASM and its RFSP-level implementation in C.}
  \label{fig-qasm-c-rfsp}
  \vspace{-2mm}
\end{figure}

Many existing quantum control systems employ custom instruction sets to program RFSP~\cite{kasprowicz2020artiq,stefanazzi2022qick,xu2023qubic}, creating fragmented software ecosystems with dedicated compilers, drivers, and libraries for each system. Despite differences in quantum-specific operations, a large portion of classical instructions---arithmetic, branching, and memory access---remains common across platforms, motivating the adoption of RISC-V as the shared base ISA.
RISC-Q generates RISC-V-compatible controllers, enabling reuse of the broader RISC-V software ecosystem, including compilers, libraries, and development tools. The resulting controllers expose a familiar bare-metal programming interface. Quantum-specific peripherals such as RF signal processors and accelerators are controlled via memory-mapped I/O or custom instructions, with their interfaces abstracted through reusable C libraries and drivers. As a result, RFSP programs can be written in high-level languages such as C, C++, or Rust, compiled to RISC-V instructions using standard toolchains, and executed directly on the controller generated by RISC-Q. 
This compatibility enables reuse of application code across different hardware implementations and quantum experiments, and lays the foundation for future operating systems for quantum hardware. 

\vspace{-2mm}
\subsection{Integration}

RISC-Q provides flexible approaches to integrate different peripherals to meet the diverse hardware requirements of quantum platforms. As an example, superconducting qubits require high carrier frequencies, whereas trapped ions or neutral atoms demand more frequency components for the RF generator. Controllers therefore need integration of a variety of peripherals including RF signal processors and custom accelerators (Fig. \ref{fig-arch}) for different experiments.

RISC-Q supports two integration methods inspired by VexiiRiscv~\cite{vexiiriscv}: \emph{memory-mapped I/O (MMIO)} and \emph{custom instructions}.

MMIO maps peripheral registers into the RISC-V processor’s address space using SpinalHDL’s TileLink library. Simple operations---e.g., configuring an RF generator’s frequency---can be performed with standard \texttt{lw}/\texttt{sw} instructions and require minimal hardware configuration.
For latency-critical paths, MMIO can introduce nondeterministic delays, whereas custom instructions provide deterministic access.
For example, QubiC~\cite{xu2023qubic} adopts 128-bit instructions to configure all parameters of a RF generators atomically.

Implementing custom instructions in pipelined processors typically requires editing files across decode, execute, and writeback stages, which is error-prone and hard to maintain. RISC-Q avoids this by using the plugin framework of VexiiRiscv and SpinalHDL’s pipeline library, which expose stage-local signals through a clean API. Each custom instruction is implemented in its own plugin with parameterized pipeline hooks, leaving the core processor codebase unmodified. This modularity improves maintainability, encourages collaborative development, and enables reuse of hardware features across projects.

\vspace{-2mm}
\subsection{Parameterization}
In addition to varying peripheral types, the internal implementation of each peripheral often differs across quantum experiments. RISC-Q supports fine-grained parameterization and interface abstraction to maximize reuse and adaptability of components.

For example, both RF signal generators and decoders require high-frequency carrier generators, but their throughput and latency requirements may differ. A parameterized carrier generator can be tailored to these differing constraints without duplicating design effort. SpinalHDL offers rich support for parameterization, including component interfaces, subcomponent implementations. This enables developers to adapt a single component design across a wide range of use cases with minimal modification.
Fig.~\ref{fig-intf} illustrates the software and hardware stack of RISC-Q. With its compatibility with open toolchains, flexible peripheral integration, and rich parameterization support, RISC-Q provides a scalable and reusable foundation for quantum control system design.

\vspace{-2mm}
\section{Prototyping a QCSoC}
\label{sec-impl}

\begin{figure}[ht]
\centering
\includegraphics[width=0.5\columnwidth]{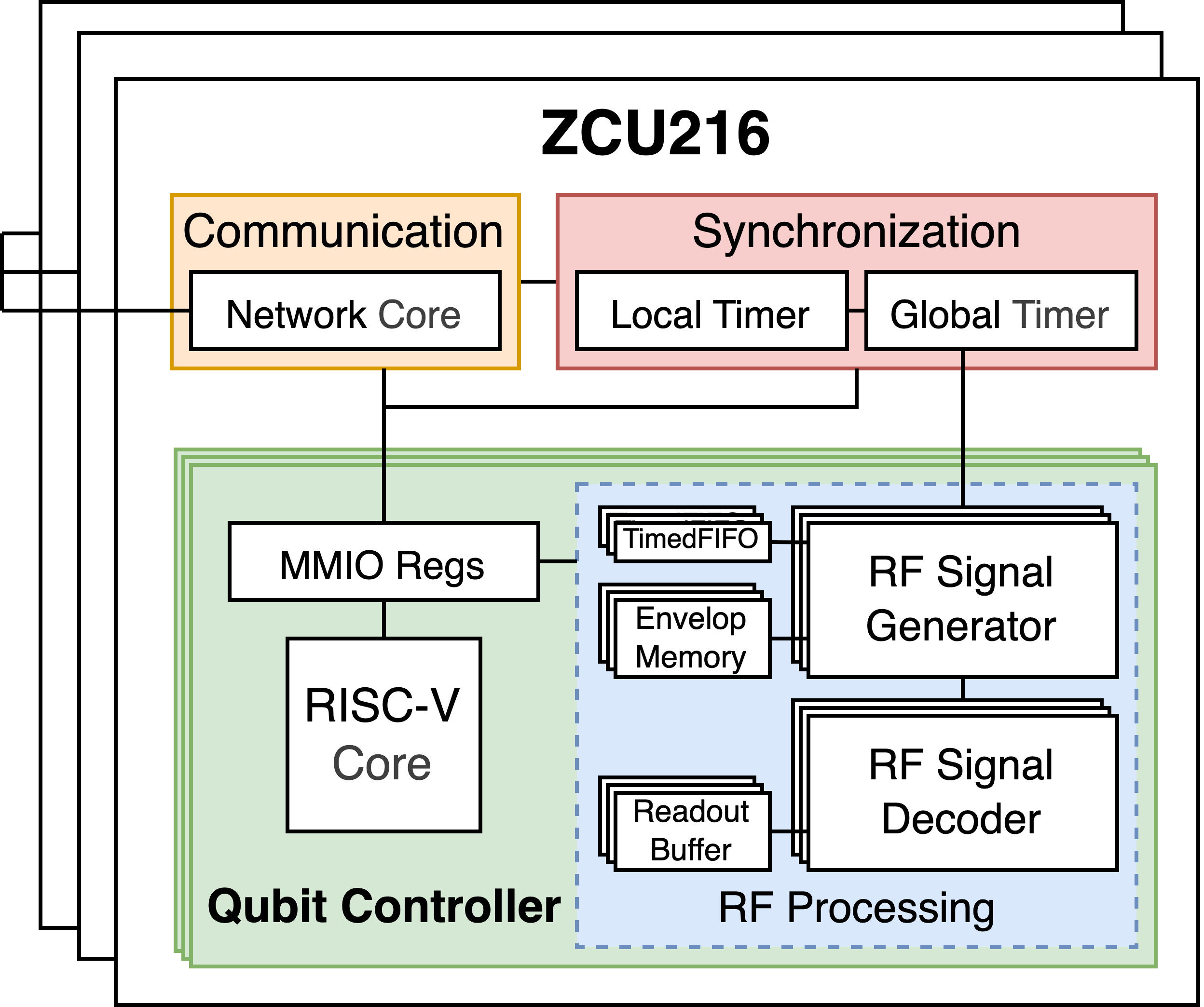}
\caption{RISC-Q-based QCSoC Architecture.}
\label{fig-qcsoc}
\end{figure}

We implement a prototype QCSoC for superconducting qubits using the RISC-Q generator. Superconducting platforms impose stringent real-time constraints on their control stack, but primarily require RF signal generation, fast readout, and low-latency feedback, rather than large domain-specific accelerators. This makes them an ideal target for validating RISC-Q’s architecture, real-time performance and scalability.
The prototype is deployed on the AMD ZCU216 RFSoC evaluation board, which integrates 16 RF DAC/ADC channels for qubit control and 4 SFP+ interfaces for fiber-optic networking. These resources are incorporated into a parameterized QCSoC design composed of RISC-V cores, RF signal processors, a low-latency network core, and a globally synchronized timer. Figure~\ref{fig-qcsoc} provides an overview of the architecture.

\vspace{-2mm}
\subsection{Qubit Controller}
Since superconducting qubits can be driven differently in parallel, we assign each qubit to a dedicated qubit controller. Each qubit controller integrates a lightweight RISC-V core tightly coupled to a parameterized RF processing subsystem capable of driving a configurable number of qubits.
Each superconducting qubit requires at least two RF signal generators: one for gate operations and one for measurement readout. Gate signals require one dedicated DAC channel per qubit, while readout signals may share a DAC channel across multiple qubits. Given the ZCU216’s 16 DAC channels, and reserving two for shared readout drives, a single board can control up to 14 qubits.

\textbf{Parameterized Qubit Assignment:}
The number of qubits assigned to each controller is fully parameterized to match different experimental requirements. Assigning one qubit per controller maximizes command throughput, as in QubiC, whereas using a single shared controller simplifies cross-qubit feedback and coordination, as done in QICK. This parameterization is realized through a SpinalHDL library that maps RF processor ports into the RISC-V core’s memory space according to the configuration.

\textbf{Signal Generation:}
The signal generation subsystem is responsible for producing high-fidelity RF pulses for gate and readout operations. RF signals are generated using Direct Digital Synthesis (DDS), where carrier samples generated in real-time are multiplied by envelope data stored in dedicated memories. The pipelined SpinalHDL implementation generates 16 samples per cycle at 500 MHz clock rate, achieving up to 8 GigaSamples Per Second (GSPS), satisfying superconducting qubit requirements.

All generator parameters including frequency, phase, amplitude, and envelope are configurable via MMIO by default. For experiments requiring rapid reconfiguration, RISC-Q additionally supports QubiC-style custom instructions that update all generator parameters in a single cycle, eliminating MMIO overhead.

\textbf{Timing Alignment:}
Accurate pulse timing is achieved through a decoupled scheduling mechanism. Timed FIFOs~\cite{stefanazzi2022qick} buffer timestamped pulse parameters and release them according to a globally synchronized timer, fully decoupling processor execution timing from pulse generation and ensuring coherent, phase-aligned control across channels.

\textbf{Readout Processing:}
Measurement is performed by RF signal decoders, which demodulate and integrate ADC data to obtain binary measurement outcomes. The integration window and demodulation parameters are programmable, enabling real-time calibration during long experiment. The decoder supports up to 8 GSPS, sufficient for the ZCU216’s 2.5 GSPS ADC throughput. Optional readout buffers capture raw ADC samples for calibration; these buffers may be disabled to allocate memory for envelopes or software.

\vspace{-2mm}
\subsection{Communication}
To meet strict real-time requirements, the QCSoC integrates lightweight network cores~\cite{eoscore} built directly on the RFSoC’s gigabit transceivers. The link layer uses a custom 64B/66B–encoded protocol with a fixed, minimal frame format. By eliminating the additional framing and handshake cycles used in AMD Aurora~\cite{amdaurora}, the custom protocol reduces one-way link latency from roughly 450 ns~\cite{xu2025multi} to about 156 ns, substantially improving responsiveness for multi-board applications such as QEC decoding.

\vspace{-2mm}
\subsection{Synchronization}
Cross-board synchronization is provided by a globally aligned timers. In our prototype, all boards share a common reference clock, minimizing drift and reducing the synchronization to the alignment of global timer registers. The offset between the local and global timer are corrected through a minimal Precision Time Protocol (PTP)~\cite{xu2025multi} implemented over the same GigaTransceiver links used for data communication. Boards exchange timestamped messages to estimate and compensate offsets, achieving sub-nanosecond alignment for coherent multi-board gate execution.

\vspace{-2mm}
\subsection{Software}
Since the controller adheres to the RISC-V ISA, it can be programmed in any language that targets RISC-V. In our prototype, control software is written in C, using MMIO to configure peripherals. Built on MMIO, a library provides a function-call interface for gate and measurement operations.
This bare-metal programming model enables control logic to be written in C, C++, Rust, and other high-level languages. Leveraging standard RISC-V toolchains, the approach supports portable and reusable control libraries and facilitates collaborative, modular software development across diverse hardware platforms.

\vspace{-2mm}
\section{Evaluation} \label{sec-eval}

\subsection{Hardware Evaluation}
\subsubsection{\textbf{Hardware Resource}}
\begin{figure}[ht]\centering
\includegraphics[width=0.618\columnwidth]{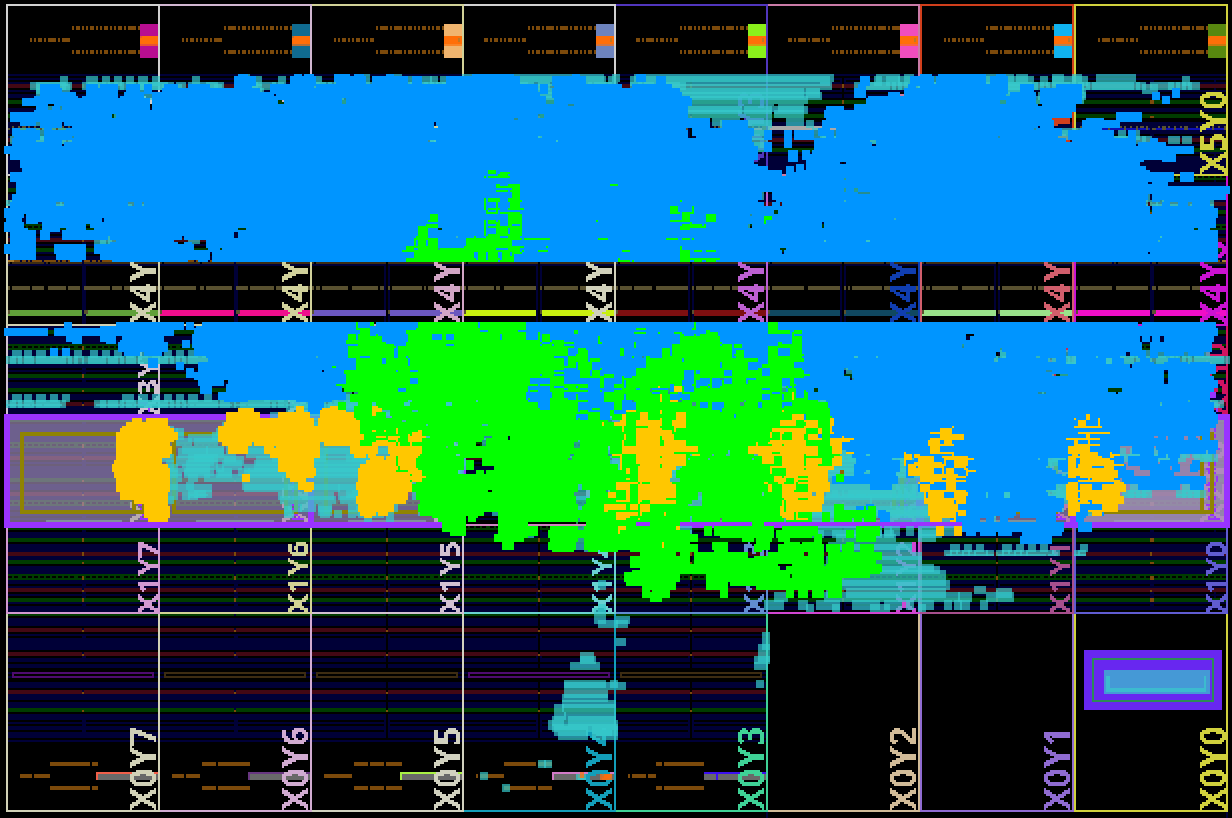}
\caption{FPGA floorplan of a 14-qubit RISC-Q QCSoC implementation on the AMD ZCU216 RFSoC. 
Orange areas correspond to RISC-V controller cores, green areas to RF signal decoders, and blue areas to RF signal generators.}
\vspace{-4mm}
\label{fig-fpga-floorplan}
\end{figure}

Table~\ref{tab-utils-freq} reports FPGA resource utilization and achievable frequency for major components of the QCSoC. Figure~\ref{fig-fpga-floorplan} shows the corresponding floorplan. The RF signal generators and decoders dominate the overall footprint, primarily due to the DDS logic. In contrast, the 14 RISC-V controller cores account for a relatively small fraction of LUTs and BRAMs, indicating that the multicore control architecture scales efficiently with qubit count.

While our evaluation focuses on FPGA prototypes, RISC-Q’s architecture is also compatible with ASIC. However, a full ASIC study must consider the mixed-signal frontend, particularly the DACs and ADCs, which largely determines signal fidelity. As this work focuses on the digital design of the QCSoC, an end-to-end ASIC analysis is left as an important future work. 

\begin{table}[ht]
  \scriptsize
  \centering
  \caption{FPGA Resource Utilization and Operating Frequency of a 14-qubit RISC-Q QCSoC on AMD ZCU216 RFSoC.}
  \label{tab-utils-freq}
  \begin{tabular}{lccccc}
    \toprule
    \textbf{Module} & \textbf{LUTs} & \textbf{FFs} & \textbf{BRAMs} & \textbf{DSPs} & \textbf{Fmax (MHz)} \\
    \midrule
    ($14 \times$) RISC-V cores & 11231 & 11509 & 28 & -- & 500 \\
    ($28 \times$) RF generators & 63649 & 156611 & 448 & 1372 & 500 \\
    ($14 \times$) RF decoders & 28382 & 95593 & -- & 700 & 500 \\
    Network core & 224 & 450 & -- & -- & 156.25 \\
    \midrule
    \textbf{Total}  & 127249 & 266372 & 476 & 1722 & -- \\
    \bottomrule
  \end{tabular}
  \vspace{-3mm}
\end{table}

\vspace{-2mm}
\subsubsection{\textbf{Performance}}
\begin{figure}[ht]
\centering
\begin{subfigure}[b]{0.6\columnwidth}
\centering
\includegraphics[width=\linewidth]{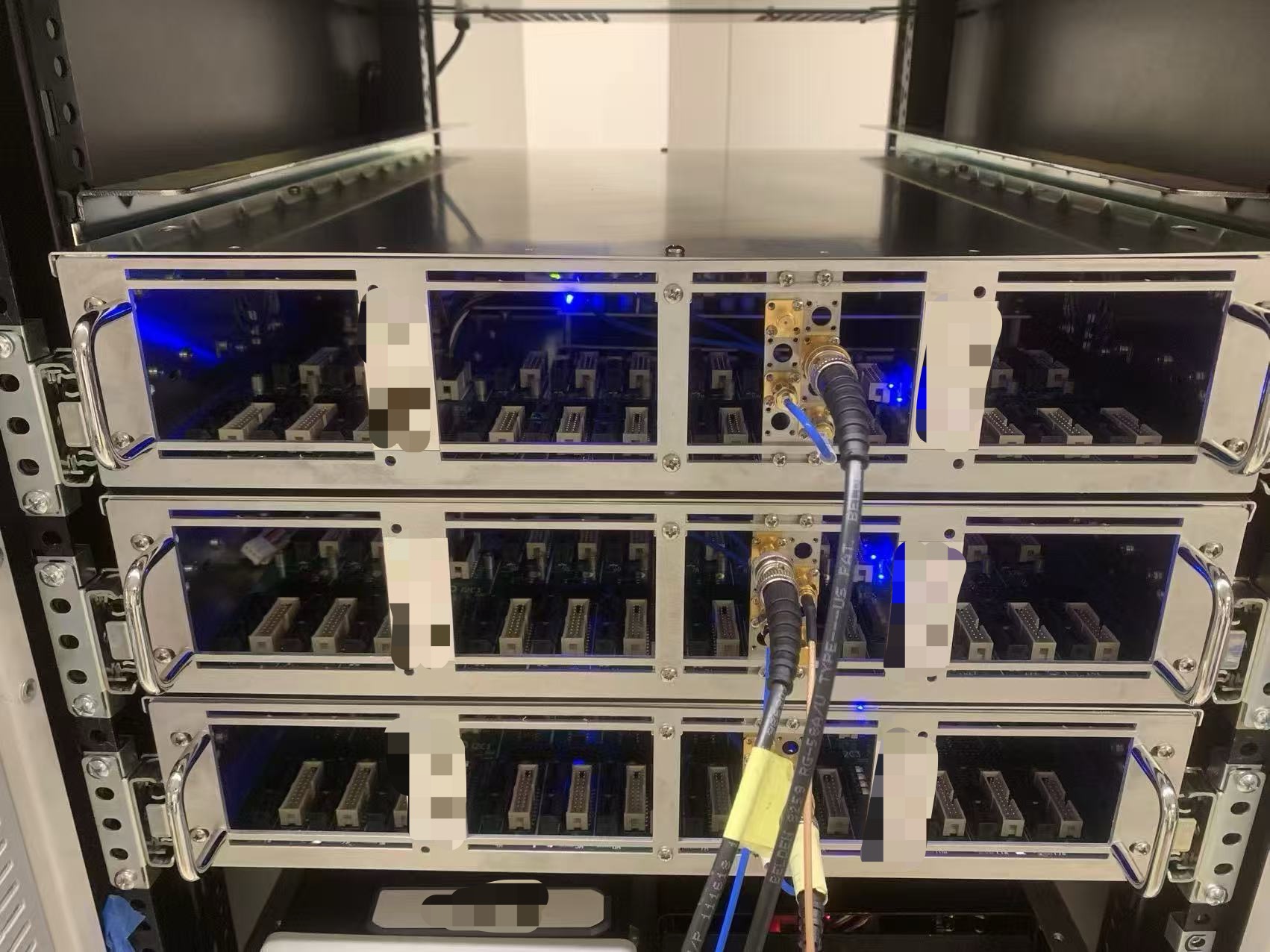}
\subcaption{}
\end{subfigure}
\begin{subfigure}[b]{0.34\columnwidth}
\centering
\includegraphics[width=\linewidth]{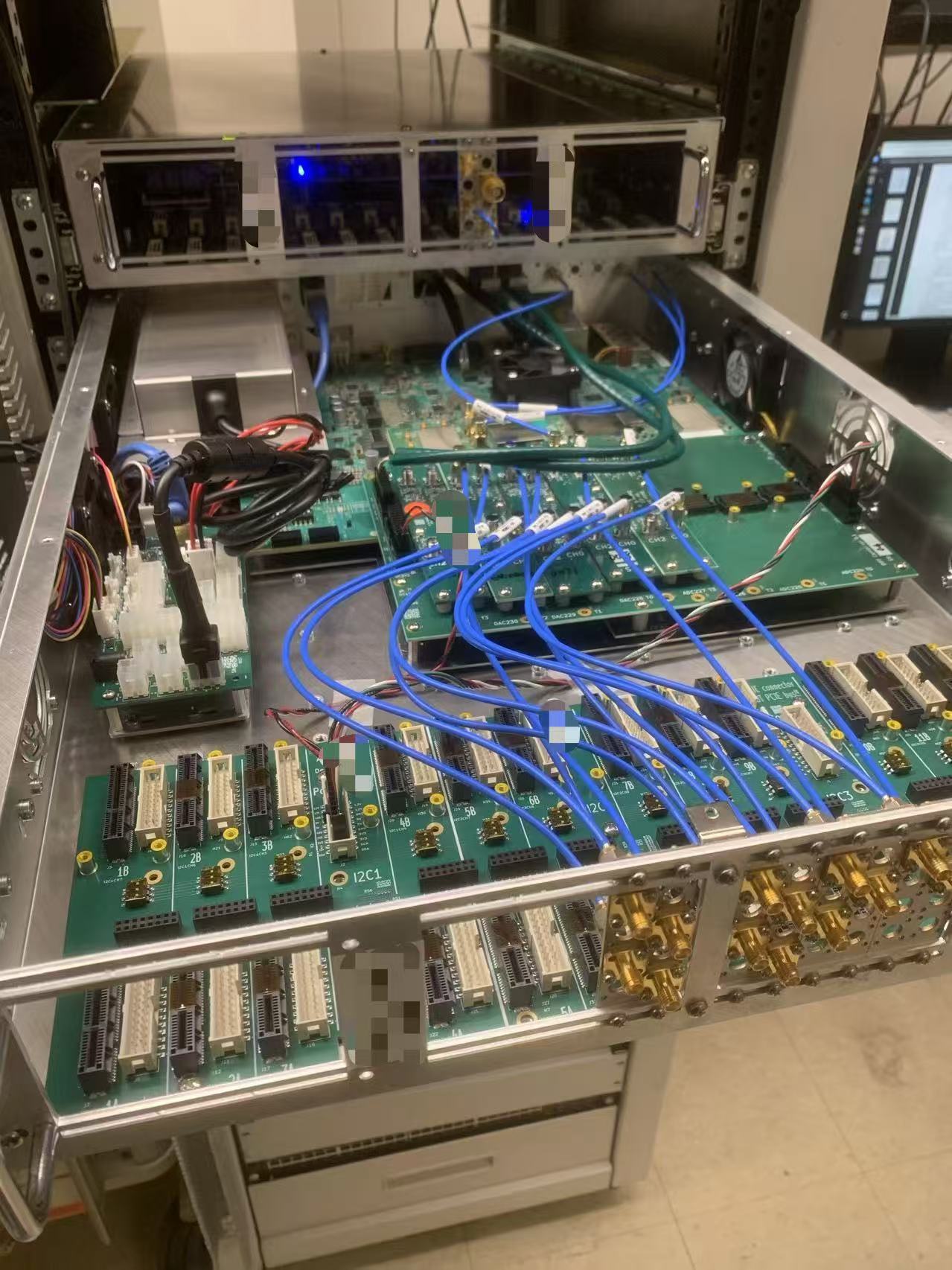}
\subcaption{}
\end{subfigure}
\caption{(a) Experimental setup using AMD ZCU216 RFSoC evaluation boards hosted in separate QubiC chassis. (b) Components inside a single chassis, including the ZCU216 board, analog front-end board, and power management board.}
\vspace{-4mm}
\label{fig-3-board-setup}
\end{figure}

Table~\ref{table-performance} summarizes key performance metrics in comparison with QICK and QubiC. All measurements are conducted on AMD ZCU216 RFSoC boards deployed in the open-source QubiC hardware environment~\cite{xu2023qubic}, shown in Figure~\ref{fig-3-board-setup}, which provides a laboratory-ready setup including analog front-end boards and regulated power delivery.
RISC-Q supports the largest number of output tones and controlled qubits per board among the three systems. Moreover, RISC-Q significantly reduces network latency relative to QubiC by replacing the AMD Aurora IP core~\cite{amdaurora} with a custom lightweight network stack \cite{eoscore} tailored for short messages.

Feedback latency is evaluated using the fast-reset operation, where a qubit is measured and driven to $\ket{0}$ by applying an $X$ pulse if the measurement outcome is 1. Latency is measured from the moment the binary measurement outcome becomes available to the generation of the feedback pulse samples, plus ADC and DAC latencies. RISC-Q achieves the lowest value (144ns), enabled by the branchless conditional-gate execution described in Section~\ref{sec-cond-gate}. 
Network latency is measured as the one-way communication delay for transmitting a 64-bit packet between two directly connected QCSoCs. This metric is directly relevant to multi-board QEC operation, where syndromes and errors must traverse the network with minimal latency (Section~\ref{sec-qec}). RISC-Q is the only system among the three with integrated support for real-time QEC.

\begin{table}[h]
  \centering
  \begin{threeparttable}
    \caption{Comparison of key specifications for three open-source quantum control systems based on AMD ZCU216.}
    \label{table-performance}
    {
    \footnotesize
    \begin{tabular*}{\linewidth}{@{\extracolsep{\fill}}cccc}
      \toprule
      & QICK \cite{stefanazzi2022qick} & QubiC \cite{xu2023qubic} & RISC-Q \\
      \midrule
      Controller FMax / MHz & 384 & 500 & 500 \\
      Output Lines / Board & 13 \cite{silva2025manarat} & 16 \cite{fruitwala2024distributed} & 16 \\
      Output Tones / Board & 20 \cite{silva2025manarat} & 16 \cite{fruitwala2024distributed} & 28 \\
      Qubits / Board & 6 \cite{silva2025manarat} & 8 \cite{fruitwala2024distributed} & 14 \\
      Feedback Latency / ns & 184 \cite{stefanazzi2022qick} & 150  \cite{hashim2025quasiprobabilistic} \tnote{a} & 144 \tnote{a} \\
      Network Latency / ns & N/A & 450 \cite{xu2025multi} \tnote{b} & 156 \tnote{b}  \\
      Jitter / ps & N/A & 1.8  \cite{xu2023qubic} \tnote{c} & 1.43 \tnote{c} \\
      QEC Integration & \ding{56} & \ding{56} & \ding{52} \\
      \bottomrule
    \end{tabular*}
    }
    \begin{tablenotes}
      \footnotesize
      \item[a] Feedback latency is evaluated in the fast reset task.
      \item[b] Network latency is evaluated for the one-way communication latency between two directly connected QCSoCs.
      \item[c] Root-Mean-Square (RMS) jitter integrating from 10 Hz to 100 MHz at 6.5 GHz.
    \end{tablenotes}
  \end{threeparttable}
  \vspace{-1mm}
\end{table}

\vspace{-1mm}
\subsubsection{\textbf{Development Effort}}
To provide an intuitive comparison, we quantify the hardware design complexity in terms of lines of HDL code. The current implementation of RISC-Q consists of 13,048 lines of HDL code, compared to 172,294 lines in QICK and 45,835 lines in QubiC.
While the codebases of QICK and QubiC include legacy modules and unused components that may inflate their line counts, RISC-Q still exhibits a substantial reduction ($\geq 75\%$) in code volume. 
More importantly, its modular, extensible architecture enables adding new peripherals without altering existing components, improving maintainability and scalability.

\subsection{Case Study}
RISC-Q’s RISC-V programming interface and modular SoC architecture enable efficient implementation of advanced quantum-control functionality. We highlight three representative examples.

\subsubsection{\textbf{Co-Design of Conditional Gates}}
\label{sec-cond-gate}

Conditional gates are essential in dynamic quantum circuits like the fast reset operation, where gate execution depends on mid-circuit measurement outcomes. Branch-based implementations, used in QICK and QubiC, incur pipeline stalls and unpredictable latency, as branch prediction is ineffective for inherently random measurement results.

RISC-Q enables an alternative branchless implementation. Both candidate gate parameters are preloaded into parallel timed FIFOs feeding the RF generator, and a control register selects the active path based on the measurement outcome:

\vspace{-1mm}
{\scriptsize
\begin{lstlisting}[language=c]
int result = *RD_RES_ADDR(7);
*MULTIPLEX_REG_ADDR(7) = result;
\end{lstlisting}
}
\vspace{-1mm}
This approach eliminates control-flow changes and minimizes latency at the cost of additional FIFOs and MMIO logic. Because RISC-Q exposes a clean RISC-V ISA and parameterized hardware components, developers can choose between branch-based and branchless variants depending on latency and resource constraints.

\subsubsection{\textbf{On-chip Calibration}}
Quantum systems are highly sensitive to environmental and internal fluctuations, requiring frequent recalibration of control parameters such as RF frequency and amplitude. Recent work~\cite{vepsalainen2022improving, gilbert2023demand} shows that lightweight closed-loop methods can maintain calibration over long experiments, but they often depend on host-controller communication, which limits feedback speed. On-chip calibration removes this bottleneck by avoiding data transfer, enabling faster correction, though implementing such protocols is difficult because current controllers must be programmed in custom low-level assembly languages.

To demonstrate the programmability of RISC-Q, we implement the amplitude calibration protocol from~\cite{gilbert2023demand} directly on the controller using the C programming language. The routine is implemented as a C function and can be easily integrated into RF signal processing programs, allowing online calibration to run between shots in high-repetition experiments. This reduces manual intervention and helps maintain high gate fidelity throughout the experimental process.

\subsubsection{\textbf{Scalable Real-Time QEC System}}
\label{sec-qec}
Quantum error correction (QEC) is essential for large-scale quantum computing because physical qubits have error rates far higher than classical bits. QEC protects quantum information by encoding a logical qubit across multiple physical qubits and detecting errors through repeated syndrome measurements, which are decoded across space and time.

A control system suitable for real-time QEC must satisfy three requirements: (1) High-throughput syndrome decoding to prevent backlog across many rounds; (2) Low-latency feedback to maintain a viable logical clock rate; and (3) Scalable architecture capable of supporting the number of qubits required for practical codes. These constraints strongly motivate hardware-centric solutions, as demonstrated by a recent real-time QEC prototype \cite{caune2024demonstrating}, where dedicated decoding accelerators and hardware-only feedback paths were essential for meeting performance goals.

Using RISC-Q, we build a distributed real-time QEC system by integrating multi-core QCSoCs across multiple AMD ZCU216 RFSoC boards with a hardware decoder~\cite{liyanage2023scalable}, connected via a low-latency fiber-optic network in a tree topology. Each QCSoC at leaf nodes performs qubit control, aggregates syndrome bits, and sends syndrome packets to the decoder at root node with a one-way latency of 156~ns. Decoded errors are returned to the leaf nodes for immediate feedback control.
We measure the decoding-feedback latency---from the final syndrome bit becoming available to the decoded correction being ready for feedback at the leaf nodes. Our implementation achieves 446~ns for a distance-3 surface code with 17 emulated qubits, an order-of-magnitude improvement over the 8.5~µs reported in the SOTA prototype~\cite{caune2024demonstrating} for an 8-qubit experiment.

A key advantage over~\cite{caune2024demonstrating} is that the RISC-Q connects controllers, decoder, and network modules with a tightly integrated datapath, avoiding the multi-layer device boundary between the Rigetti control system and the Riverlane QEC decoder. Using RISC-Q’s generator-based methodology, the entire multi-board QEC system is implemented with only $\sim$3000 lines of additional HDL, demonstrating both scalability and rapid hardware development.

\section{Conclusion}
We present RISC-Q, an open-source and flexible generator for QCSoCs with RISC-V compatibility. RISC-Q enables the reproduction of a fully functional QCSoC for superconducting qubit systems with significantly reduced development effort, while exceeding the state-of-the-art performance. Its support for fine-grained parameterization and flexible peripheral integration facilitates efficient design space exploration and rapid prototyping. We believe RISC-Q could significantly enhance productivity in the development of quantum control, micro-architecture, and low-level operating systems, which are crucial infrastructures for enabling practical, scalable, and high-performance quantum computing.

RISC-Q can benefit the entire quantum computing community --experimental physicists, application designers, quantum control developers, and system researchers alike. By offering a unified, extensible platform, it fosters collaboration and shared innovation in quantum control system design. In this work, we demonstrate a RISC-Q prototype for superconducting qubits and plan to extend it to neutral-atom and trapped-ion platforms in the near future.
\newpage
\balance
\bibliographystyle{ACM-Reference-Format}
\bibliography{riscq}

\end{document}